# Self-ordering induces multiple topological transitions for elastic waves in phononic crystals

Jiujiu Chen,[1,*] Hongbo Huang,[1] Shaoyong Huo,[1] Zhuhua Tan,[2,†] Xiaoping Xie,[1] and Jianchun Cheng[3,‡]

[1]*State Key Laboratory of Advanced Design and Manufacturing for Vehicle Body, College of Mechanical and Vehicle Engineering, Hunan University, Changsha 410082, China*

[2]*School of Mechanical Engineering, Hebei University of Technology, Tianjin, 300401, China*

[3]*Department of Physics, Nanjing University, Nanjing 210093, China*

(Dated: April 23, 2018)

**Abstract:** Topological defects with symmetry-breaking phase transitions have captured much attention. Vortex generated by topological defects exhibits exotic properties and its flow direction can be switched by altering the spin configurations. Contrary to electromagnetic and acoustic domains, the topological transport of elastic waves in periodic structures with topological defects is not well explored due to the mode conversion between the longitudinal and transverse modes. Here, we propose an elastic topological insulator with spontaneously broken symmetry based on the topological theory of defects and homotopy theory. Multiple topological transitions for elastic waves are achieved by topologically modifying the ellipse orientation in a triangular lattice of elliptical cylinders. The solid system, independent of the number of molecules in order parameter space, breaks through the limit of the point-group symmetry to emulate elastic pseudospin-orbit coupling. The transport robustness of the edge states is experimentally demonstrated. Our approach provides new possibilities for controlling and transporting elastic waves.

Topological defects in ordered media [1] with spontaneously broken symmetry have attracted an enormous interest due to their nontrivial topology, which can play a central role in physical processes such as phase transitions [2-4]. Their topological origin and fundamental behavior was first described by the Kibble-Zurek mechanism as a continuous system is quenched across a phase transition into an ordered state [5,6]. In recent years, such topological defects have been extensively studied in various branches of physics. It has been shown that the topological defects with +1/2 or -1/2 topological charge can govern cell motion [7-9], and even arise in fatigue of materials [10]. Antivortices and vortices can be generated in three-dimensional nonporous ferroelectric structures with topological defects [11]. These exotic physical phenomena and unprecedented material properties imply that topological defects can be leveraged to explore quantum behavior of classical waves and new forms of topological orders in condensed matter systems.

The discovery of topologically ordered states is an exciting field of research inspired by quantum systems. These concepts have lately been extended to classical systems, including electromagnetic [12-25], acoustic [26-40], and elastic waves [41-48] systems. Given the nature of Fermi electrons is fundamentally different from bosonic phonons, there are two ways to realize topologically protected wave propagation in phononic media. One possible way is to emulate the quantum Hall effect [27-31,42,43] by breaking the time reversal symmetry. Another alternative is to employ solely passive components to establish analogues of the quantum valley Hall effect [37-39] and the quantum spin Hall effect [32-35,48]. Recently, an acoustic analogue of the quantum spin Hall phase transition has been theoretically and experimentally achieved by appropriately tuning the parameters or folding the Brillouin zone [32-35]. Compared with acoustic domains that only support longitudinal polarization, experimentally realizing the topological phase in elastic solids has so far remained difficult owing to the existence of both transverse and longitudinal polarizations [42]. Moreover, these systems have high modal densities and large acoustic impedance mismatch, making them vulnerable to structure defects and enhancing strong backscattering at the boundaries between distinct materials. Thus, these challenges have hindered the realization of topological protection for in-plane bulk elastic waves on an integrated platform.

Here, we introduce the concept of topological defects to solid phononic crystals (PCs) with spontaneously broken symmetry. This system consists of a triangular lattice of elliptical tungsten cylinders embedded in an epoxy resin [Fig. 1(a)]. The lattice constant is $a_s = 15$ mm and the distance between the center of each cluster to the centroid of each ellipse is $l = a_s/3$. The major axis of the ellipse is $a = 1.644$ mm and the minor axis is $b = 1.370$ mm. In the following, the theory of topological defects [1-3] is adopted to describe the elementary topology of the PCs by rotating individual ellipses in a unit cell to different directions. As schematically shown in Fig. 1(b), the orientational modification can be mapped on the order parameter space $\theta$, which is defined as the angle between the major axis of an ellipse and the given *x*-axis, and it is written as [1]

$$\theta = k\varphi + C, \tag{1}$$

where $\varphi = tan^{-1}(y/x)$ represents the polar angle of the center position of the ellipse, *C* is the initial phase for $\varphi = 0$. The integer *k* is the topological charge and is equivalent to the winding number *n*. It is calculated by measuring the total rotation in $\theta$-space and can be defined by $n = \oint_{\delta\Omega} \nabla\theta \cdot dL$, where $\delta\Omega$ denotes a closed contour and $L(l,\varphi)$ denotes the polar coordinates. Note that the topological modification structure is characterized by the winding number and the initial phase, in which the winding number is used to classify different homotopy classes while the initial phase is used to distinguish between the modification structures in each homotopy class. Owing to the symmetry of $\theta$-space in our system, $n$ can take on the values of +1, 0. Herein, $n = +1$ with respect to vortex and hedgehog structure, and $n = 0$ with respect to trivial rectilinear modification structures. The ellipse-shaped cylinders are anisotropic and the use of the initial phase enables the system to exhibit different symmetries. Thus, these structures as a topological modification in the orientational order markedly change the topological properties of the solid PCs and can be used to emulate the elastic pseudospin-orbit coupling.

According to the topological theory of defects and homotopy theory [1-3], the $n = +1$ structure is characterized by the nature of topologically stability. To reveal the characteristic of the topological modification structure with $n = +1$ and create an elastic topological insulator for in-plane bulk elastic waves, we demonstrate that the spatially inhomogeneous variation of the ellipse orientation can pose enhanced control over topological phase, causing energy band inversion. Figure 1(c) shows the eigenfrequencies spectra of *p*-type and *d*-type states at the Brillouin zone center evolving for a sweep of the initial phase *C*. It is observed that the doubly degenerated band-edge frequencies vary smoothly and the eigenstates exchange occurs at $C = 47.5°$ and $C = 132.5°$, resulting in a double Dirac cone with a fourfold accidental degeneracy at the Brillouin zone center. With inherent time reversal invariants, the rotation of the crystalline domains introduces chirality to the structures with $C = 47.5°$ and $C = 132.5°$, implying that the corresponding band structures are exactly the same, as illustrated in Fig. 1(d). It is shown that the solid PCs studied here have a broken continuous symmetry and undergo a symmetry inversion in reciprocal space, which ultimately induces multiple topological phase transitions.

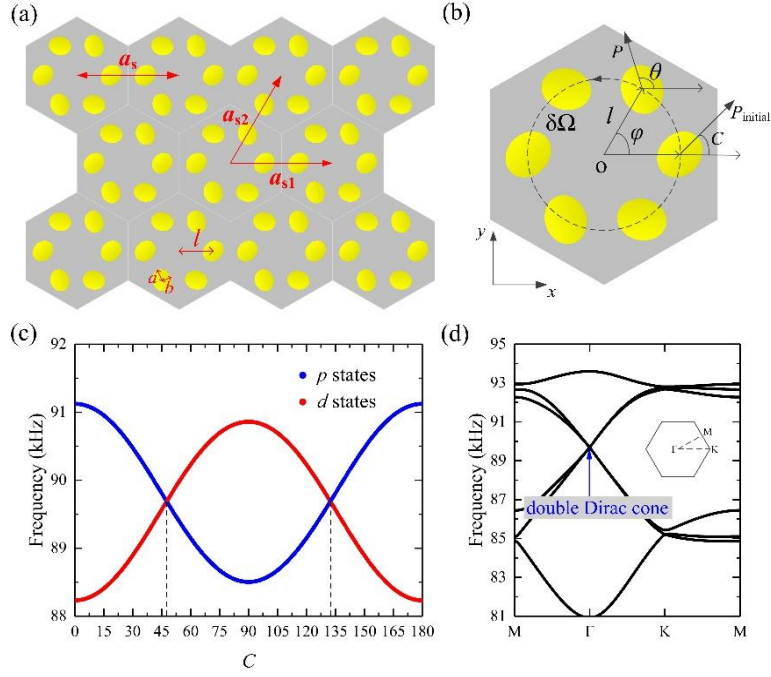

FIG. 1. (a) Schematic illustration of a triangular-lattice of hexagonal clusters composed by six elliptical cylinders embedded in an epoxy resin. (b) Unit cell diagram and order parameter space of the PCs. (c) Eigenfrequencies of the $p$- and $d$- type states at the $\Gamma$ point evolving for a sweep of the initial phase $C$. (d) Calculated band structure in which $C = 47.5°$ and $C = 132.5°$, showing the fourfold degenerated double Dirac cone at the $\Gamma$ point. (Inset: Brillouin zone of triangular lattice).

We then introduce a perturbation to emulate elastic pseudospin-orbit coupling, accompanied by the opening of a topological band gap at the $\Gamma$ point. To this end, the fourfold degeneracy needs to be lifted and split into two twofold degenerate states. Figures 2(a) and 2(b) show the band structures of the radically configuration with $C = 0°$ and the azimuthally configuration with $C = 90°$, respectively. Seen in the insets are the associated displacement field distributions of the degenerate elastic eigenstates at the Brillouin zone center, which respectively corresponding to the colored dots at the $\Gamma$ point. The elastic spin-1/2 states can thus be realized through hybridizing the $p/d$ states as $p_{\pm} = (p_1 \pm ip_2)/\sqrt{2}$ and $d_{\pm} = (d_1 \pm id_2)/\sqrt{2}$, which is protected by a pseudo time-reversal operator $T$ ($T^2 = -1$). For $C = 0°$, the two bands below the gap are of $d$-type, whereas the two bands above are $p$-type as per the analogy to electronic orbital shapes. For $C = 90°$, the band gap still exists, however the position of the $p$ and $d$ is now inverted, i.e., $p$-type states are at the lower frequency compared to $d$-type states. Such a band inversion signifies a topological phase transition that occurs as the gap reopens, which is characterized by a nonzero $Z_2$ topological invariant. To determine the topology of the band gaps, we employ the $k \cdot P$ perturbation method [32,33,35] to construct an effective Hamiltonian and further calculate the spin Chern numbers (see Note 1 of Supplemental Materials [49]). As expected, we obtain the spin Chern numbers $C_s = \pm 1$ in the case of the radically configuration, indicating that the band gap in Fig. 2(a) is nontrivial in topology. Differently, the azimuthally configuration has the spin Chern numbers $C_s = 0$, which signifies a topologically trivial band gap in Fig. 2(b). These results further confirm the appearance of a phase transition from nonzero Chern numbers to zero Chern numbers in two structures. Considering the broad generality in practical systems, we demonstrate that the topological modification in the orientational order is also applicable to other classic waves, such as solid-in-air and air-in-fluid acoustic systems (see Note 2 of Supplemental Materials [49]). We also discuss the trivial rectilinear modification structures with a winding number corresponding to $n = 0$ without band inversion (see Note 3 of Supplemental Materials [49]). In particular, as long as the order parameter maintains its rigidity and the scatterer has $C_2$ symmetry, the phase transition induced by specific topological modification cannot be removed by smooth, local deformations (see Fig. S4 of Supplemental Materials [49]).

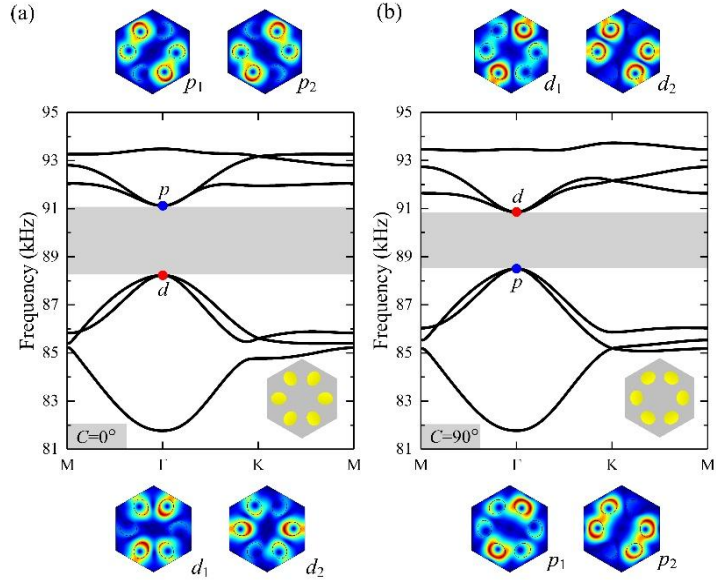

FIG. 2. Band structures of the solid PCs with different initial phase. Unit cells are depicted at the bottom right corner in each subfigure. Panel (a) represents the radically configuration with the initial phase $C = 0°$ and the associated displacement field distributions for the degenerated eigenstates at the Γ point. Panel (b) represents the azimuthally configuration with the initial phase $C = 90°$, the p- and d-type states are inverted with respect to the band gap which indicates the occurrence of band inversion. The band gap (gray shaded region) in (a) is topologically nontrivial, whereas the one in (b) corresponds to topologically

trivial regime.

To ascertain the appearance of topologically protected helical edge states for in-plane bulk elastic waves in our system, we combine the radically configuration ($C = 0°$) and the azimuthally configuration ($C = 90°$) to construct a domain wall (highlighted with a dashed red line) [see Fig. 3(a)]. Figure 3(b) shows the simulated projected band structure of a supercell consisting of total 20 unit cells in the *y* direction. As expected, a pair of gapped edge states (shown in blue and red curves) exist evidently inside the overlapped bulk band gap of the two PCs. The two elastic edges states have opposite group velocities at a given frequency, implying the existence of pseudospin-orbit coupling and counter-propagation of edge states. In Fig. 3(c), we plot the simulated field maps of the elastic displacement of edge states corresponding to points A and B, respectively, and both maps show that A and B states are localized at the interface between the nontrivial and trivial phase. Further, the time-averaged mechanical energy flux ($I_j = -\sigma_{ij} v_j$, where $\sigma_{ij}$ and $v_j$ are stress tensor and velocity vector, respectively) and their directions (indicated by the black arrows) confirm the existence of the pseudospin-up state (rotates anti-clockwise) and the pseudospin-down state (rotates clockwise), which unveils the spin nature of these in-plane modes in the solid PCs. Remark that, there is a tiny gap at the Γ point, arising from the breaking of the symmetry at the domain wall between the trivial and nontrivial phase. However, these edge modes are still topological due to the distinct topology of the two surrounding bulks and the topological edge propagation remains robust.

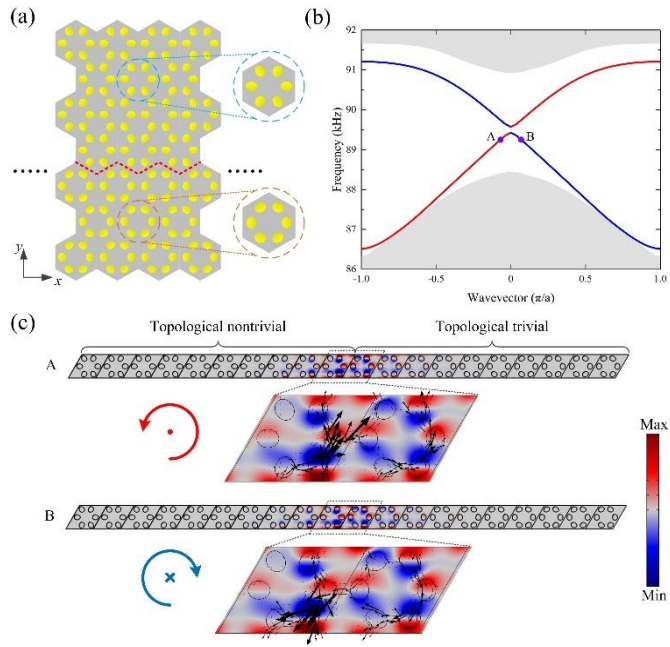

FIG. 3. (a) Schematic of elastic topologically-protected domain wall, in which radically configuration is replicated at the upper domain wall and azimuthally configuration is replicated at lower domain wall, respectively. The domain wall is marked by the dashed red line. (b) Calculated band structure for a 1 by 20 supercell with a domain wall at center. The blue and red lines represent edge states of opposite group velocities, while the shaded regions indicate the bulk states. (c) The corresponding displacement field distribution of A and B states. Black arrows indicate the time-averaged mechanical energy flux.

The solid system is independent of the molecular numbers in $\theta$-space whose topological phase can be specified by the polar angle $\varphi$, which allows for an extension of phononic topological phase down to a variety of topological modification structure based on different $\theta$-spaces. To verify this, we investigate the allotropes of a hexagonal non-bravais lattice containing twelve elliptical tungsten cylinders embedded in an epoxy resin. The lattice constant is $a_s = 15$ mm and $l = 0.3695a_s$. The major axis of the ellipse is $a = 1.2$ mm and the minor axis is $b = 0.6$ mm. By introducing a topological modification in the unit cell, multiple topological phase transitions can occur (see Fig. S5(a) of Supplemental Materials [49]). Figures 4(a)-(c) show the band structures of the solid PCs with different initial phase. A fourfold degenerated double Dirac cone appears at the Γ point of the structure with $C = 22°$. As shown in Figs. 4(a) and 4(c), the *p*- and *d*- type states are inverted between two structures, causing energy band inversion. The band gap closing and reopening through Dirac points indicate that the solid PCs experience a topological transition from a topological trivial crystal $(C = 0°)$ to a topological nontrivial crystal $(C = 45°)$. In Fig. 4(d), we show that there are two edge states of opposite velocity in the overlapped bulk band gap when connecting the two PCs of different band topology. The corresponding energy flow profiles at points A and B further confirm the existence of these topologically protected helical edge states. The associated displacement field maps and spatial amplitude profile of the edge states in the *y* direction are plotted in Fig. 4(e), and shows that the displacement field is concentrated along the domain wall and decays exponentially into the bulk with respect to $y = 0$, which is the key characteristic of edge states. We have also numerically investigated a topologically protected one-way waveguide with bends to examine the topological robustness against defects, and no notable scattering losses are observed at the domain wall (see Fig. S5(b) of Supplemental Materials [49]). Note that the topological phase transition can be realized in a double Dirac cone of PCs beyond $C_{6v}$ symmetry (see Fig. S6 of Supplemental Materials [49]). The overall results hence confirm the high flexibility of this topological modification without the limit of the point-group symmetry and the lattice geometry, which clearly distinguishes itself from the conventional methods with the unparalleled advances in practical applications. It is worth emphasizing that, our current research is focused on a specific type of topological charge $(k=1)$, such mapping $\theta = k\varphi + C$ can realize diversified topological orders of phononic states with distinct characteristic by tuning the values of $k$ and $\varphi$. Therefore, a tremendous order parameter space remains to be explored.

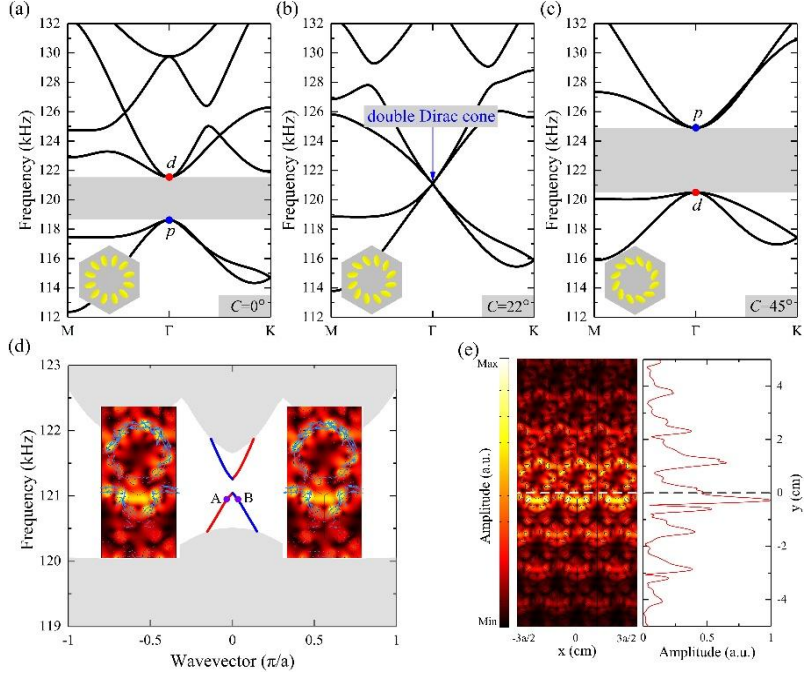

FIG. 4. (a)-(c) Band structures of the solid PCs with different initial phase with $C = 0°$, $C = 22°$ and $C = 45°$, respectively. Unit cells are depicted at the bottom left corner in each subfigure. (d) Calculated band structure of a supercell consisting of a topological nontrivial crystal $(C = 45°)$ stacked with a topological trivial crystal $(C = 0°)$. The blue and red lines represent an elastic spin+ and spin- edge state, respectively. Insert shows the corresponding energy flux profiles for the edge modes, indicated by the blue arrows. The shaded regions indicate the bulk states. (e) Spatial amplitude profile of the displacement fields localized at the domain wall.

One of the most striking features of the helical edge states is the backscattering-immune one-way propagation and the topological robustness against defects. In the following, we will move on to an experimental demonstration of these observations. Considering the broad generality of the proposal in elastic solid and facilitating the processing and manufacturing of experimental samples, we have investigated the topological properties of in-plane bulk elastic waves in two-dimensional solid PCs constituted of a triangular array of the elliptical steel cylinders embedded in an epoxy resin matrix, obtaining similar results (see Fig. S7 of Supplemental Materials [49]), and have further demonstrated experimentally the topological wave propagation. Fig. 5(a) shows a picture of the fabricated samples of the elastic topologically-protected waveguide (see Note 5 of Supplemental Materials [49]). The simulated displacement field distribution of the topologically-protected straight waveguide and the ordinary phononic crystal waveguide at a bulk band gap frequency (125.1 kHz), is illustrated in Figs. 5(b)-(c) to verify the reflection-free transmission of the helical edge states. It is worth noting that, in comparison with the trivial system alone, in which pure isolation occurs, the in-plane bulk elastic waves transmit along the domain wall without backscattering for the straight waveguide. As shown in Fig. 5(e), the experimental results illustrate the measured ~ 20 dB transmission enhancement of the topological waveguide in contrast to the ordinary waveguide. To further test the robustness of these edge states, we introduce a series of sharp bends to the topologically protected waveguide as shown in Fig. 5(d) and it is observed that there is no notable scattering losses at the structural bends due to the topological protection. The experimentally measured transmission spectra are presented in Fig. 5(e) and demonstrate no significant elastic transmission drops within the bulk band gap frequency region by comparing with the results of the straight waveguide. Thus our edge states exhibit topological protection against certain disorder and defects, demonstrating the absence of backscattering and resulting in the inherently robust one-way propagation of in-plane bulk elastic waves.

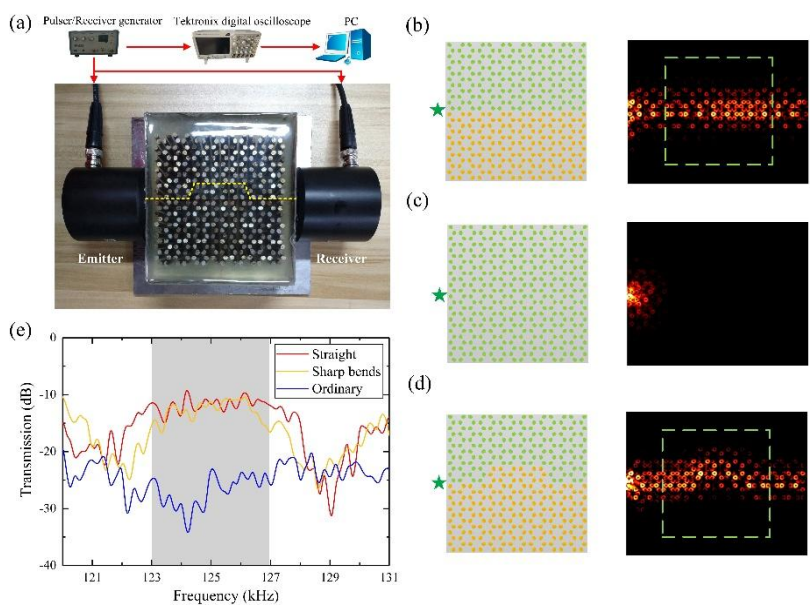

FIG. 5. (a) Photograph of the experiment setup. The yellow dashed lines indicate location of the waveguides constructed by stacking a topological trivial crystal $(C = 90°)$ with a topological nontrivial crystal $(C = 0°)$. (b)-(d) Simulated displacement field distribution of the topologically protected waveguide and the ordinary waveguide at a bulk band gap frequency (125.1 kHz). The green dashed rectangle denotes the region of the area 125 × 110 mm$^2$ measured in experiments. Green and orange regions represent the trivial and nontrivial crystals. (e) Experimentally measured transmission spectra for these elastic topologically protected waveguide and ordinary waveguide. The shaded region corresponds to the topological band gap.

In conclusion, we have designed a unique elastic topological insulator based on a topological modification in a triangular lattice of elliptical cylinders, whose multiple topological phase transitions can be realized by inhomogeneously changing the ellipse orientation within the unit cells. Robust one-way propagation for in-plane bulk elastic waves and the ability to guide waves along channels with sharp bends in the solid PCs is experimentally verified. Interestingly, the solid system with spontaneously broken symmetry is independent of the number of molecules in order parameter space, providing the possibility for achieving a new form of topological wave propagation beyond $C_{6v}$ symmetry. Such topological modification is much different from conventional methods to obtain topological phononic phase in condensed matter physics, which allows us to explore fundamentally new physics beyond the original ones via various order parameter spaces. Moreover, the methods can be directly extended to other classic wave such as electromagnetic and plate-mode waves. We believe that our demonstrated elastic topological phenomena open a new avenue for manipulating and transporting elastic waves, which can be of immense value for practical applications.

This work is supported by the National Science Foundation of China (Grant No.11374093 and No.11672214); Young Scholar fund sponsored by common university and college of the province in Hunan.

J.-J.C. and H.-B.H. contributed equally to this work.

*jjchen@hnu.edu.cn; †zhtan@hnu.edu.cn; ‡jccheng@nju.edu.cn